\documentclass[12pt,a4paper]{article}

\usepackage{a4wide}
\usepackage{graphicx}
\usepackage{amsmath}
\usepackage{amssymb}
\usepackage{xspace}

\newcommand{\GeV}{\ensuremath{\,\mathrm{GeV}}}

\newcommand{\Dzero}{D\O\xspace}
\newcommand{\order}[1]{{\cal O}\left(#1\right)}

\title{A note on the CDF high-$p_t$ charged particle excess}

\author{Matteo Cacciari,$^{1,2}$ Gavin~P.~Salam$^1$ and Matthew~J.~Strassler$^3$\\
  {\it  \normalsize $^1$LPTHE, UPMC Univ.~Paris 6 and CNRS UMR 7589, Paris, France}\\
  {\it  \normalsize $^2$Universit\'e Paris Diderot, Paris, France}\\
  {\it  \normalsize $^3$NHETC, Rutgers University, New Jersey, USA}
}

\date{}

\begin{document}
\maketitle
\begin{abstract}
  It has recently been pointed out that CDF data for the cross
  section of high-$p_t$ charged particles show an excess of up to
  three orders of magnitude over QCD predictions, a feature
  tentatively ascribed to possible violations of
  factorisation.
  We observe that for $p_t > 80\GeV$ the measured charged-particle
  cross sections become of the same order as jet cross sections.
  Combining this information with data on charged particle
  distributions within jets allows us to rule out the hypothesis that
  the CDF data could be interpreted in terms of
  QCD factorisation violation.
  We also comment on the difficulty of interpreting the excess in
  terms of new physics scenarios.
\end{abstract}

In reference \cite{Aaltonen:2009ne}, the CDF collaboration at the
Tevatron collider has published measurements of the inclusive charged
particle spectrum at transverse momenta, $p_t$,  up to $150\GeV$.
These data are based on a minimum-bias trigger sample consisting of about
16\,000\,000 events and a smaller high track-multiplicity trigger sample.
In its analysis, CDF compared the data to Monte Carlo predictions for
a somewhat larger minimum-bias sample from the Pythia event generator
\cite{Sjostrand:2006za} and found that approaching $50\GeV$ the data
exceeded predictions by about a factor of three.
The comparison was not extended beyond $50\GeV$.

\begin{figure}
  \centering
  \includegraphics[width=0.45\textwidth]{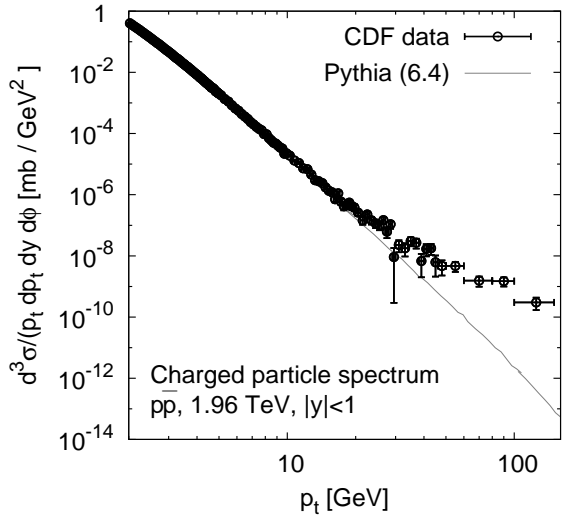}
  \caption{Comparison of the CDF charged-particle
    data~\cite{Aaltonen:2009ne} with predictions from Pythia
    6.421~\cite{Sjostrand:2006za} using the DW
    tune~\cite{Albrow:2006rt}. }
  \label{fig:charged+pythia}
\end{figure}

Recently, Albino, Kniehl and Kramer \cite{Albino:2010em} presented a
comparison of the CDF data to a NLO QCD calculation involving parton
distribution functions and fragmentation functions from global fits.
Such a calculation should be more precise than predictions from
Pythia and should have better constrained uncertainties.
Furthermore, it covers the full extent of the high-$p_t$ region in
which there are data.
The striking observation in \cite{Albino:2010em} is
that in the highest $p_t$ bin, $100-150\GeV$, the data exceed QCD
predictions by a factor of up to $\order{10^3}$.
Independent work by Arleo, d'Enterria and Yoon \cite{Arleo:2010kw},
which appeared while our note was being finalised, made the same
observation and also compared to a Pythia
sample that covered the whole $p_t$ range,
confirming the findings of a large excess.
This excess is illustrated in fig.~\ref{fig:charged+pythia}, where we
compare the data to the combination of a Pythia minimum-bias sample in
the low-$p_t$ range with dijet samples at high $p_t$.

In interpreting their findings, the authors of
ref.~\cite{Albino:2010em} argue that the comparison of data and theory
challenges the validity of QCD factorisation,
requiring a drastic modification either to the theoretical calculation
or the experimental data.
We can investigate the question of factorisation in more
detail by comparing the charged-particle data to Tevatron data on
inclusive jet production.

\begin{figure}[p]
  \centering
  \includegraphics[width=0.56\textwidth]{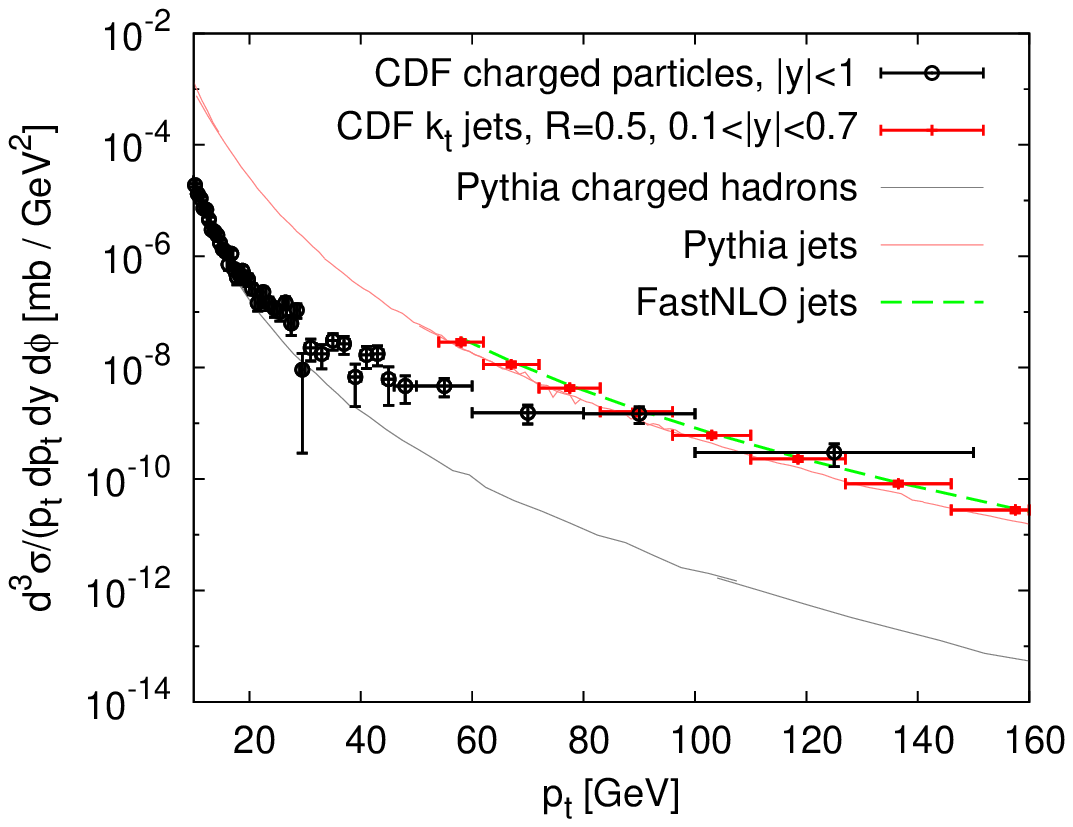}\hfill
  \includegraphics[width=0.42\textwidth]{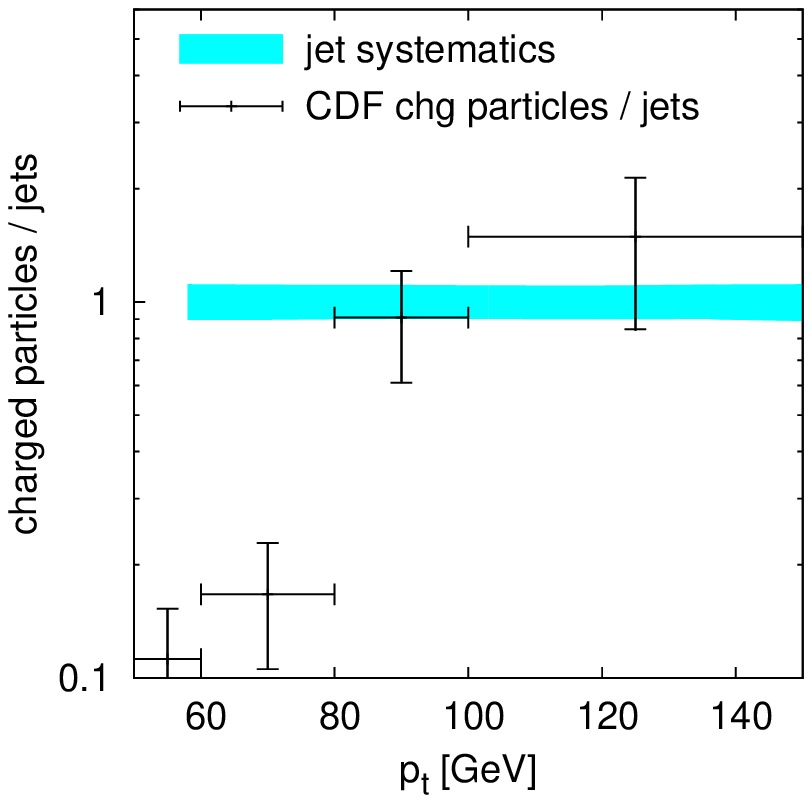}
  \caption{Left: comparison of the charged-particle
    data~\cite{Aaltonen:2009ne} with CDF data on the inclusive jet
    spectrum \cite{Abulencia:2007ez}, showing also predictions from
    Pythia and the NLO calculation for the jets from
    FastNLO and NLOJet\texttt{++} ~\cite{Kluge:2006xs,NLOJet} with  CTEQ66
    PDFs~\cite{Nadolsky:2008zw}.
    Right: ratio of the charged-particle spectrum to the (rebinned)
    CDF inclusive jet spectrum. 
    Note that the charged-particle and jets data correspond to
    slightly different rapidity ranges.  For the $p_t$ range of
    relevance, the mismatch in rapidity ranges implies only modest
    additional corrections, $\order{10\%}$ (which have not been
    applied).  }
  \label{fig:charged+jets}
\end{figure}

The particle and jet spectra are compared in
fig.~\ref{fig:charged+jets}. The left-hand plot shows the inclusive
charged particle data \cite{Aaltonen:2009ne} together with data from
CDF for the inclusive jet spectrum~\cite{Abulencia:2007ez} (with the
$k_t$ algorithm~\cite{KtHH}, using $R=0.5$; results with other jet
definitions and from \Dzero~\cite{Abazov:2008hua} are similar).
The right-hand plot shows the ratio of the charged particle to jet
spectra.
One sees that for $p_t > 80\GeV$ the charged particle cross sections
become of the same order as the jet cross sections.
If the CDF charged particles are normal hadrons, then the only
explanation for the high-$p_t$ bins of fig.~\ref{fig:charged+jets}
would be that for the majority of jets, the whole jet momentum is
accounted for by a single charged hadron.

\begin{figure}[p]
  \centering
  \includegraphics[width=0.5\textwidth]{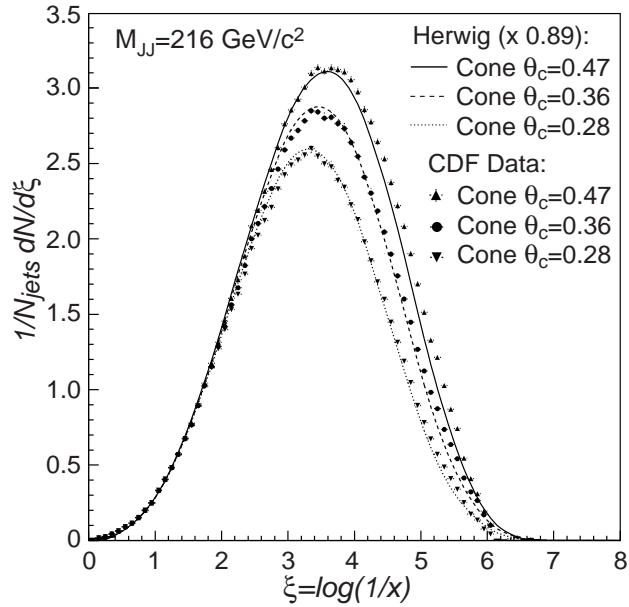}
  \caption{Figure~20 of ref.~\cite{Acosta:2002gg}, by the CDF
    Collaboration, showing the inclusive distribution of momentum
    fraction $x$ of charged particles in cones around each of the two
    jets axes in dijet events at the Tevatron
    (Run~I).}
  \label{fig:cdf-fig20}
\end{figure}
Such a feature would not just be a violation of factorisation for
fragmentation functions, but would imply that above some parton
energy, collinear splitting, the mechanism by which a parton's
momentum is shared among multiple daughters in a parton shower, would
turn off.
Aside from violating fundamental properties of quantum field theory,
an interpretation of this kind is in contradiction with CDF data
on charged particle spectra within jets, such as
fig.~\ref{fig:cdf-fig20}, taken from
ref.~\cite{Acosta:2002gg}:
for a dijet mass range of $200-260\GeV$, corresponding to $p_{t,jet}
\sim 100\GeV$, those data show that only about $0.1\%$ of jets contain
a charged hadron carrying at least $90\%$ of the jet
momentum.\footnote{
  Ref.~\cite{Acosta:2002gg} does not measure all particles inside the
  jet, but only those within a cone of limited radius. 
  However, any particle carrying a large fraction of the jet's
  momentum has to be near the centre of the jet. Furthermore,
  measurements of jet shapes~\cite{Acosta:2005ix} constrain the
  fraction of energy that can be in the outer parts of a jet (beyond
  an angle of 0.5 radians) to an average of $10-15\%$.
  A second restriction of \cite{Acosta:2002gg} is that it only
  considers events in which the two hardest jets are well-balanced in
  $p_t$ and any 3rd or 4th jet is much softer than the two leading
  ones. 
  While this does not affect conclusions about the validity of QCD
  fragmentation within jets, it can have implications for alternative
  explanations of the origin of the excess.}

The argument can be further refined by noting that the jet spectrum
$d\sigma/dp_t$ falls as $p_{t}^{-n}$ with $n\simeq6$ in the relevant
$p_t$ range and that the measured inclusive charged particle
momentum-fraction distribution within the jet, $C(x)$, does not depend
too strongly on the jet $p_t$.
By taking the $(n-1)^\mathrm{th}$ moment of $C(x)$, $\int_0^1 dx\,
x^{n-1} C(x)$, as determined from the
points in fig.~\ref{fig:cdf-fig20}, one then derives an
expectation for the ratio of charged-particle and jet inclusive
spectra. Numerically this turns out to be of order $0.006$ with
uncertainties of a few tens of percent (related to the exact choice of
$n$ and the finite size of the bins of the data for $C(x)$).
Such a result is consistent with the predictions of parton
shower programs like Pythia (cf.\ the two Pythia curves in
fig.~\ref{fig:charged+jets} left).
This should not be a surprise: the data of \cite{Acosta:2002gg} had
been compared to Herwig \cite{Herwig} simulations, showing near
perfect agreement.

We therefore conclude that the large excess in charged-particle
production at high transverse momenta cannot be explained by any
modification of QCD fragmentation functions (or of other aspects of QCD
factorisation), because this would lead to a strong contradiction with
Tevatron data on charged tracks within jets in the same $p_t$ range.%
\footnote{ 
  Ref.~\cite{Arleo:2010kw} similarly reaches the conclusion
  that a breakdown of factorisation could not be responsible for the
  observed excess. Its reasoning is based on the observation that the
  scaling of $d\sigma/dp_t^2$ with $p_t$ and with centre of mass
  energy $\sqrt{s}$ (at fixed $p_t/\sqrt{s}$) is anomalously close to
  the $1/s^2$ and $1/p_{t}^4$ powers that arise based from pure
  dimensional arguments without scaling violations.
  While we are sympathetic to such an argument, we believe that ours
  involves fewer theoretical assumptions.
}

The most likely explanation for the CDF excess, as argued already
in~\cite{Arleo:2010kw}, is probably some issue with the data.
But it is worth considering what the implications would be if the
tracks are real.

Unless somehow jets in multijet events fragment in a way so different
from jets in dijet events that the plots of \cite{Acosta:2002gg} are not
applicable in multijet QCD, then the high $p_T$ tracks in
\cite{Aaltonen:2009ne} cannot form part of an ordinary jet -- and so,
if the tracks are real, they cannot be ordinary hadrons.  Nor can they
be electrons or muons; non-isolated electrons would appear as jets,
while such high rates for muons, converted photons 
or isolated electrons are inconsistent
with other data.  Could a new metastable exotic object be responsible
for these tracks?  This seems unlikely but is not completely trivial
to rule out.

Whatever these objects are, they must be
produced with a very large cross-section. From the last 4 bins of the
data, one sees that it must be of order 50
nanobarns, with large errors, if the track-objects are pair produced.

This big cross section is a major impediment to a new physics
explanation. 
A colour octet fermion with mass of 25 GeV and $p_t > 50\GeV$ would be pair
produced with $\sim 10$ nb cross-sections at
leading order.
A wide resonance with a small branching fraction to dijets and a large
branching fraction to exotica, such as a 150 GeV ``coloron'' in the
spirit of \cite{JHUcoloron}, could have a few-nanobarn cross-section
through mixing with the gluon, and a decay dominantly to an exotic
final state, while drowning in dijet background.
Still, even though a 50~nb cross-section pushes the limits of credulity, it
seems worthwhile to try to exclude any exotica directly
from searches for those objects.

For an object to leave a track, it must be ionising: it must carry
electric charge or magnetic charge on an atomic scale.  (It could be a
neutral object with a large dipole moment.)
In addition, to not affect the dijet fragmentation results in
\cite{Acosta:2002gg}, it must not appear frequently in dijet events,
or else must deposit only a fraction of its $p_T$ as calorimetric jet
energy (so that it appears as a small contribution at negative $\xi$
in lower dijet-mass bins).
A possible candidate, massive metastable charged particles (``CHAMPs''
or ``CMSPs''), if colourless and stable, are excluded by CDF
\cite{CDFCHAMPS} and \Dzero \cite{D0CMSP}.%
\footnote{A fractionally charged CHAMP (still
allowed by cosmology and direct searches if the CHAMP pair are quirks
\cite{luty} that will always eventually find each other and
annihilate) would have its track $p_T$ overestimated in
\cite{Aaltonen:2009ne}, but the powerful CHAMP/CMSP bounds of
\cite{CDFCHAMPS,D0CMSP} still exclude it.  Any decrease in efficiency
in the muon system with the lower charge would be partly countered by
a decreasing tracking efficiency, requiring an even larger
cross-section.} %
Even non-hadronic CHAMPS that decay in flight are excluded: given
how many must leave tracks, too many would still reach the muon system
and be detected by CDF's CHAMP search.
(The \Dzero search, requiring two ``muons'' per
event, is less sensitive.)

Charged R-hadrons (bound states of an exotic coloured object $Q$ with
quarks and gluons) are more subtle.  Understanding of their behaviour
in matter has continued to evolve (for a review see
\cite{SMPreview}.)  It certainly seems unlikely that a few million
R-hadrons could escape the CDF CHAMP search, but it is not simple because
of charge-flipping, by which an R-hadron colliding with a nucleus may
shift its charge by 1 unit.  Such flipping reduces dramatically the
number of good ``muon'' candidates produced by R-hadrons, and
makes the \Dzero search, with its requirement of two ``muons'', one with
tight isolation, far less sensitive.  The CDF study
directly considers stable top squarks, and, with some assumptions
regarding their properties in matter, reports a detection efficiency
greater than $3.5\%$.  But loopholes might remain for lower-mass
R-hadrons, for example those built from a charge-neutral colour-octet
$Q$,
which is only
constrained at LEP \cite{ALEPHgluino,DELPHIgluino} to be above about
25 GeV.  (Ref.~\cite{KaplanSchwartz} argues that indirect mass limits
can be placed at 50 GeV.) 
And colour octets, less well
understood than triplets, may well have a lower detection efficiency
\cite{SMPreview}.

There are good reasons to expect substantial missing transverse
momentum (MET) from R-hadrons~\cite{SMPreview}. 
One is that slow R-hadrons have much less kinetic energy than $p_T$.
Another is that fast R-hadrons that are neutral (if any are produced)
are expected to deposit only a fraction of their kinetic energy in the
calorimeter.
In association with initial state radiation (ISR), a substantial
MET signal would then be generated.
Indeed in \cite{Hewett} it is argued that a robust lower bound of 170
GeV can be placed on gluino masses: specifically, neutral R-hadrons
leave too little energy to register as jets, and therefore contribute
to monojet signals \cite{D0monojet,CDFmonojet}.
But R-hadron energy deposition increases at low masses, so perhaps
this argument might break down for sufficiently light R-hadrons.
Moreover, even accepting the arguments of
\cite{Hewett}, R-hadrons decaying in flight with $c\tau\gamma \sim
1-2$ meters might still be allowed; such decays would further
reduce the sensitivity of the CHAMP/CMSP searches.
Still, it would be necessary either that the R-hadrons appear
typically in events with more than two jets, or that their decay
include invisible particles, so as to evade the constraints from
\cite{Acosta:2002gg} discussed above and not distort the jet spectrum.

We certainly think the R-hadron scenario unlikely.  And while there
may well be other allowed exotic objects that could give tracks of low
curvature without producing corresponding high-$p_T$
jets,\footnote{For instance, in some regimes a quirk bound state ---
  an electric dipole formed from two charged particles bound together
  by a hidden-sector flux tube --- could be misread as a high-$p_T$
  track \cite{luty} and yet, because of unusual behaviour in matter,
  might not be easily detected by existing CHAMP/CMSP searches.
  However it seems very difficult to achieve the required
  cross-sections.} a plausible new physics scenario seems difficult to
achieve.  But because of the range of possibilities and the
uncertainties on their dynamics, we cannot claim to have strictly
eliminated this option.

We therefore suggest that CDF, with appropriate scepticism, might wish
to exclude such possibilities directly from the data, if no technical
problem with the data is immediately apparent.  It would seem
that any exotic explanation for the tracks is likely to become
evident in a direct perusal of the signal events with the high-$p_T$
tracks in~\cite{Aaltonen:2009ne}.  Signatures could include a
systematic mismatch of a track's $p_T$ with its associated energy
deposition, unusual energy deposition patterns, time-of-flight delay
or large $dE/dx$, evidence of a late decay, overall calorimeter MET,
pairs of tracks per event, etc.  (By contrast a fake track would not
lead to calorimetric MET, and two such tracks per event would be very
rare.  Note that misreconstructed tracks with unphysically high
momentum are removed from the CDF CHAMP search \cite{CDFCHAMPS},
though no details on this step are given.)  Should a direct
investigation of the existing events prove inconclusive, it is likely
that the event sample can be enlarged by looking in other trigger
streams, given the signal's high rate and apparent high energy.  At a
minimum, ISR radiation should contribute events of this type to
(prescaled) 20 or 50 GeV jet triggers.

To conclude, we have argued that the excess of high-$p_t$ tracks seen
in \cite{Aaltonen:2009ne} cannot be attributed to QCD factorisation
violations, or to any other misunderstanding in QCD calculations of
hadron production.
The argument is largely based on data from the Tevatron itself, in the
same $p_t$ range as the excess.
We have not been able to eliminate fully the possibility that the excess is
due to exotic new physics, though we view it as unlikely.
Inspection of the data should be able to settle the question.

\section*{Acknowledgements}
MJS would like to thank the LPTHE and the FRIF for hospitality while
this work was carried out and C.~Kilic, A.~Lath and E.~Halkiadakis for
conversations.
MC and GPS gratefully acknowledge the French ANR for support under
contract ANR-09-BLAN-0060 and the work of MJS was supported by NSF
grant PHY-0904069 and by DOE grant DE-FG02-96ER40959.


\end{document}